\documentclass[fleqn,11pt]{article}
\usepackage{amsmath}
\usepackage{amsfonts}
\usepackage{graphicx,color}
\usepackage{hyperref}
\usepackage{epstopdf}
\hypersetup{
  colorlinks=TRUE,
  linkcolor=blue,
  urlcolor=blue,
  citecolor=blue
}
\usepackage{natbib}
\usepackage{rotate,epsfig}
 \usepackage{lscape}
\DeclareMathSizes{1}{5}{5}{5}

    {\par \noindent {\bf Proof:}}%
    {\par \indent}

\usepackage{epsfig}
\usepackage{amssymb}
\usepackage{fancyhdr}
\usepackage{amsmath}

    {\par \noindent {\bf Algorithm I:}}%
    {\par \indent}

\begin{document}

\title{Genomic Data Analysis using a Two Stage Expectation Propagation Algorithm for Analysis of Sparse Bayesian High-Dimensional Instrumental Variables Regression}
\author{.}
\author{Morteza Amini\footnote{E-mail address: morteza.amini@ut.ac.ir}\\
{\small Department of Statistics, School of
Mathematics, Statistics and Computer Science,}\\
{\small College of Science, University of Tehran,
P.O. Box 14155-6455, Tehran,Iran}}
\maketitle

\begin{abstract}
Simultaneous analysis of gene expression data and genetic variants is highly of interest, especially when the number of gene
expressions and genetic variants are both greater than the sample size. Association of both causal genes and effective SNPs
makes the use of sparse modeling of such genetic data sets, highly important. The high-dimensional sparse instrumental variables
models are one of such useful association models, which models the simultaneous relation of the gene expressions and genetic
variants with complex traits. From a Bayesian viewpoint, the sparsity can be favored using sparsity-enforcing priors such as
spike-and-slab priors. A two-stage modification of the expectation propagation (EP) algorithm is proposed and examined for
approximate inference in high-dimensional sparse instrumental variables models with spike-and-slab priors. This method is
an adoption of the classical two-stage least squares method, to be used { with the Bayes context}. A simulation
study is performed to examine the performance of the methods. The proposed method is applied to analysis of the mouse obesity data.
\end{abstract}

\noindent{\bf Keywords:} {Causal inference, Expectation propagation, Spike-and-Slab prior, Sparse instrumental variables model}

\section{Introduction}

Detection of simultaneous association of gene expressions and single nucleotide polymorphism (SNP) with complex traits,
such as obesity, heart disease and cancer, is one of the highly important issues in genome-wide studies
\citep{emilsson2008genetics}.
One of the useful models which provides a practical approach for jointly modeling the effects of genes and genetic markers
on the trait as the response, is the instrumental variables (IV) method. IV models are extensively studied in econometrics literature
\citep{MR0483259}
and observational epidemiology and causal inference (see e.g.
\citet{lawlor2008mendelian},
for a review until 2008).
The high-dimensional sparse IV models are of interest in situations in which the number of covariates and instrumental variables
are both greater than the sample size and there are too many zero coefficients (non-effective covariates and instrumental variables)
in the model. \citet{gautier2011high}
proposed a Dantzig-type variable selection method for high-dimensional IV models. \citet{MR3001131}
used the LASSO method
\citep{MR2815776}
for the first-stage covariates in a high-dimensional IV model.
Recently, \citet{MR3338502}
have proposed a two-stage regularization method, by imposing the $L_1$ penalties to both coefficients
of the covariates and instrumental variables in a high-dimensional IV model.

From a Bayesian viewpoint, the sparsity can be favored using
sparsity-enforcing priors for the model coefficients. Recently, the
sparse Bayesian models are widely applied in gene association
studies for prediction and classification
\citep[e.g.,][]{davies2017sparse, yang2017bayesian}.
The
sparsity-enforcing priors are priors which are peaked at zero or
have a large mass at zero. Laplace \citep{MR2417254},
Student's $t$
\citep{tipping2003fast},
horseshoe
\citep{carvalho2009handling}
and
spike-and-slab \citep{MR997578,MR1425430,george1997approaches}
priors are some of the most important sparsity-enforcing priors.
Among the aforementioned priors, the spike-and-slab priors are of a
special interest, partly because of their mixture structure which
allows to discriminate zero and non-zero coefficients, their
closed-form convolution with Gaussian density, which makes the
Gaussian approximation of the posteriors straightforward, and the
less shrinkage effect on the non-zero coefficients induced by the
spike-and-slab prior compared with the other priors.

Using the spike-and-slab priors, the posterior distribution can not be often computed algebraically and the
approximation methods should be used to estimate the parameters of the model. Different {asymptotically exact and}
approximate Bayesian inference are applied to sparse Bayesian models, such as Gibbs sampling
\citep{george1997approaches,hernandez2015expectation}
variational Bayes \citep{attias1999inferring,MR2896713}
and
expectation propagation (EP) algorithm \citep{MR2452620,hernandez2015expectation}.
The EP
algorithm \citep{MR2717007}
have many advantages over Gibbs sampling and variational Bayes,
including less computational cost compared to Gibbs sampling and decreasing the probability
of approximating local modes of the posterior compared to variational Bayes. Recently,
\citet{hernandez2015expectation}
have proposed an EP method for linear regression models
with spike-and-slab priors by splitting the posterior distribution into only three separate factors and
approximating them separately. They have shown that the proposed method have a low computational cost and
high precision with respect to other methods.

In this paper, we propose a two-stage modification of the EP method
to Bayesian sparse high-dimensional IV models, with spike-and-slab
prior. This proposed modification is based on the standard idea of
replacing covariates $X$ by their expectations conditional on the
instruments, as in the classical two-stage least squares (2SLS)
method \citep{anderson2005origins},
 in which the covariates $X$ are
first regressed on the instruments $Z$ and the response is then
regressed on the first stage predictors. A simulation study is conducted to examine the
performance of the proposed method. We focus on the application of the
proposed method to genetical genomic to identify potentially causal
genes as covariates and genetic variants as instrumental variables.

The rest of this paper is organized as follows. Section 2 introduces
the Bayesian modeling of the sparse instrumental variable model with
spike-and-slab priors. The proposed two-stage modification of the EP
algorithm is introduced and implemented to sparse IV model in
Section 3. The numerical illustration including the simulation study
and an analysis of the mouse obesity data is presented in Section 4,
based on the proposed method. {The details of the algorithm are given in the Appendix and the R functions to implement the proposed methods as
well as other 2-stage sparse frequentist competitors are available
at \href{https://github.com/mortamini/2Stage-Sparse-IVR}{https://github.com/mortamini/2Stage-Sparse-IVR}.}

\section{Materials and methods}

\subsection{The mouse obesity data set}

Our aim is to analyze the
mouse obesity data-set described by \citet{wang2006genetic}.
The data-set includes an F2 intercross of 334 mice derived from the
inbred strains C57BL/6J and C3H/HeJ on an apolipo-protein E (ApoE)
null background, which were fed a high-fat Western diet from 8 to 24
weeks of age. The mice were genotyped using 1327 SNPs at an average
density of 1.5 cM across the whole genome, and the gene expressions
of the liver tissues of these mice were profiled on micro-arrays
that include probes for 23,388 genes. Data on several
obesity-related clinical traits were also collected on the animals.
The genotype, gene expression, clinical data and the annotation
table of genes are available for download, respectively, at
\href{http://www.genetics.org/cgi/content/full/genetics.110.116087/DC1}{Supplementary
material of \citet{van2010expression}
},
\href{ftp://ftp.ncbi.nlm.nih.gov/pub/geo/DATA/SeriesMatrix/GSE2814/}{National
Center for Biotechnology Information Web site},
\href{http://labs.genetics.ucla.edu/horvath/CoexpressionNetwork/MouseWeight/}{Horvath's
Web page} and
\href{https://www.ncbi.nlm.nih.gov/geo/query/acc.cgi?acc=GSE2814}{the
GSE2814 information page}.

\subsection{Sparse IV model with spike-and-slab priors}

{ Suppose that $(y_i,X_i,Z_i),\; i=1,\ldots,n$, is a sample of size $n$ of scalar response variable $y$,
$1\times p$ covariate vector $X$ (e.g. gene expressions), and $1\times q$ vector of instrumental variables $Z$ (e.g. genotypes).} Consider the following IV model
\begin{equation}\label{iv}
\hspace{5cm}\begin{array}{c}
y_i=X_i\beta+\epsilon_i,\\
X_i=Z_i\Gamma+\varepsilon_i,\\
\end{array}
\end{equation}
for $i=1,\ldots,n$, where $\beta$ is a $p\times 1$ vector of unknown linear effects of the covariates, $\Gamma$ is a $q\times p$ matrix
of unknown linear effects of instrumental variables on the covariates, $\epsilon_i$ and $\varepsilon_i$ are $1\times 1$ and $1\times p$
vectors of random errors.

In order to consider the sparse high-dimensional IV model, we assume
that both $p$ and $q$ are greater than $n$ and a large subset of
coefficients in the vector $\beta$ and the matrix $\Gamma$ are zero.
{ As considered by
\citet{lopes2014bayesian},
we assume
that $(\epsilon_1:\varepsilon_1),\ldots,(\epsilon_n:\varepsilon_n)$
are independent and identically distributed from $(p+1)$-variate
normal distribution with a zero vector mean and a
variance-covariance matrix $\Sigma$.}

{From  \eqref{iv}, we can write for $i=1,\ldots,n$
\begin{align}
(y_i,X_i)&=(Z_i\Gamma\beta:Z_i\Gamma)+(\epsilon_i+\varepsilon_i\beta:\varepsilon_i)\nonumber\\
&=(Z_i\Gamma\beta:Z_i\Gamma)+(u_i:\varepsilon_i)\label{join}
\end{align}
 Thus
$$(u_i,\varepsilon_i)\stackrel{\rm iid}{\sim} N_{p+1}(0,\Omega(\beta)),$$
where $\Omega(\beta)=B\Sigma B'$ and
$$ B=\left(\begin{array}{cc}
1 & 0_{1\times p} \\
\beta & I_{p\times p}
\end{array}\right).$$}

Hence, the likelihood function of $\beta$ and $\Gamma$ is
\begin{equation}\label{like}
L(\beta,\Gamma|X,Z,y)\propto \prod_{i=1}^n\exp\left\{\frac{-1}{2}\left((y_i,X_i)-(Z_i\Gamma\beta,Z_i\Gamma)\right)\Omega(\beta)^{-1}
\left((y_i,X_i)-(Z_i\Gamma\beta,Z_i\Gamma)\right)'\right\},
\end{equation}
{where $X$ is the $n\times p$ matrix of covariates, $Z$ is the $n\times q$ vector of instruments and $y$ is a $n\times 1$ vector of responses. }

To enforce the sparsity to the parameters $\beta$ and $\Gamma$, we consider the spike-and-slab priors
\citep{MR997578,MR1425430,george1997approaches},
which are mixtures of a normal density and a point
probability mass at zero, as follows
\begin{equation}\label{pbet}
p(\beta|\eta)={\prod_{j=1}^p\left[{\cal N}(\beta_j;0,\nu_0)^{\eta_j}\delta(\beta_j)^{1-\eta_j}\right]}=\prod_{j=1}^p\left[\eta_j{\cal N}(\beta_j;0,\nu_0)+(1-\eta_j)\delta(\beta_j)\right],
\end{equation}
\begin{equation}\label{pgam}
p(\Gamma|\theta)={\prod_{j=1}^{pq}\left[{\cal N}(\gamma_j;0,\omega_0)^{\theta_j}\delta(\gamma_j)^{1-\theta_j}\right]}=\prod_{j=1}^{pq}\left[\theta_j{\cal N}(\gamma_j;0,\omega_0)+(1-\theta_j)\delta(\gamma_j)\right],
\end{equation}
where {${\cal N}(x;\mu,\sigma^2)$ stands for the probability density function of the normal
distribution with mean $\mu$ and variance $\sigma^2$, } the hyper-parameters $\eta_1,\ldots,\eta_p$ and
$\theta_1,\ldots,\theta_{pq}$ take the values 0 (for zero coefficients) and 1 (for non-zero coefficients), $\Gamma=((\gamma_{i,j}))$ is
vectorised as $\gamma=(\gamma_1,\ldots,\gamma_{pq})'$, { that is $\gamma={\rm vec}(\Gamma)$ is formed by combining the rows of $\Gamma$ end to end.} $\nu_0>0$ and $\omega_0>0$ are known
variances and $\delta(\cdot)$ is the Dirac delta function, $\delta(x)=1$, if $x=0$ and $\delta(x)=0$, otherwise.

To develop a hierarchical Bayesian analysis, the priors for the hyper-parameter
$\eta_1,\ldots,\eta_p$ and $\theta_1,\ldots,\theta_{pq}$ are considered to be Bernoulli as follows
\begin{equation}\label{peta}
p(\eta)=\prod_{j=1}^p{\rm Ber}(\eta_j;p_0),
\end{equation}
\begin{equation}\label{ptet}
p(\theta)=\prod_{j=1}^{pq}{\rm Ber}(\theta_j;\pi_0),
\end{equation}
where $p_0$ and $\pi_0$ are known prior probabilities. { These parameters are the main parameters for controlling the sparsity of the model and act like the penalty parameters in the frequentist penalized sparse model proposed by \citet{MR3338502}.}

Given $X$, $y$ and $Z$, the posterior of $\beta$, $\Gamma$, $\eta$ and $\theta$ is given by
\begin{equation}\label{post}
p(\beta,\Gamma,\eta,\theta|X,y,Z)=\frac{L(\beta,\Gamma|X,Z,y)p(\beta|\eta)p(\Gamma|\theta)p(\eta)p(\theta)}{p(y,X|Z)},
\end{equation}
where $p(y,X|Z)=\sum_{\theta=0}^1\sum_{\eta=0}^1\int \int L(\beta,\Gamma|X,Z,y)p(\beta|\eta)p(\Gamma|\theta)p(\eta)p(\theta)\; {\rm d}\beta\; {\rm d}\Gamma$.

For a given new vector of $(x^{\rm new},z^{\rm new})$, the predictive density of the response at a point $y^{\rm new}$ is computed as follows
$$p(y^{\rm new}|X,y,Z)=\sum_{\eta=0}^1\sum_{\theta=0}^1\int\int p(y^{\rm new}|x^{\rm new},\beta)p(x^{\rm new}|z^{\rm new},\Gamma)p(\beta,\Gamma,\eta,\theta|X,y,Z)\; {\rm d}\beta\; {\rm d}\Gamma,$$
where the notations $\sum_{\eta=0}^1$ and $\sum_{\theta=0}^1$ stand for
$\sum_{\eta_1=0}^1\cdots\sum_{\eta_p=0}^1$ and $\sum_{\theta_1=0}^1\cdots\sum_{\theta_{pq}=0}^1$, respectively.

\subsection{A two stage modification of the EP algorithm}

The { Bayes approximation method EP }
\citep{MR2717007}
is an algorithm
for approximation of the joint distribution of the parameters and the observed data with a simple distribution $\tilde{Q}$.

Let the likelihood function be $p(x|\theta)$ with prior $p(\theta|\eta)$ and the hyper-prior $p(\eta)$. The joint distribution of $(x,\theta,\eta)$ would be then

\begin{equation}\label{joint}
P(x,\theta,\eta) = p(x|\theta)p(\theta|\eta)p(\eta) = \prod_{i=1}^k f_i(\theta,\eta)
=Q(\theta,\eta),
\end{equation}
for a given number of factors $k$.
The aim of the EP algorithm is to approximate the components of the joint density $P(x,\theta,\eta)$,
by $\tilde{f}_1,\ldots,\tilde{f}_k$, respectively.  Each update step of the EP algorithm refines the
 parameters of $\tilde{f}_i$, $i=1,\ldots,k$, so that the Kullback-Leibler (KL) divergence between the
un-normalized distributions $f_i\tilde{Q}^{(-i)}$ and $\tilde{f}_i\tilde{Q}^{(-i)}$ is minimum, which is proved
to have a single global solution \citep{MR2247587},
where
$$\tilde{Q}^{(-i)}(\theta,\eta) =\prod_{j\neq i} \tilde{f}_j(\theta,\eta),$$
and the KL divergence between $f$ and $g$ is
$${\rm KL}(f || g) =\int f(z) \log\left(\frac{f(z)}{g(z)}\right)\;{\rm d}\mu(z),$$
for the sigma-finite measure, $\mu$.

Thus the EP algorithm is as follows:
\begin{description}
\item 1. Initialize parameters and hyper-parameters of $Q$ and approximated factors, such that all priors and hyper-priors are noninformative.
\item 2. Repeat until the parameters of $f_1,\ldots,f_k$ converge:
\begin{description}
\item 2.1. For $i=1,\ldots,k$, select $\tilde{f}_i$ to be refined.
\begin{description}
\item 2.1.1. Compute $\tilde{Q}^{(-i)}$,
\item 2.1.2. Update $\tilde{f}_i$ so that ${\rm KL}(f_i\tilde{Q}^{(-i)}|| \tilde{f}_i\tilde{Q}^{(-i)}) $ is minimized.
\end{description}
\end{description}
\end{description}
For the exponential family of distributions, the updated parameters of $\tilde{f}_i$ in step 2-1-2 are
found by matching the sufficient statistics of $f_i\tilde{Q}^{(-i)}$ and $\tilde{f}_i\tilde{Q}^{(-i)}$
\citep{MR2717007}.
Since the EP algorithm is not guaranteed to
converge in general
\citep{MR2717007},
 it can be improved by damping the update operations of EP \citep{minka2002expectation},
in step $t+1$, $t\geq 1$ of the EP algorithm, by replacing
$\tilde{f}_i^{(t+1)}$ by $(\tilde{f}_i^{(t+1)})^{\epsilon_t}(\tilde{f}_i^{(t)})^{1-\epsilon_t}$,
where the damping parameter sequence $\epsilon_t \in (0,1)$ is suggested to be a decreasing sequence, staring from a value near 1.

For Bayesian analysis of the sparse IV model \eqref{iv} using the EP algorithm, first, we have to factorize the
joint distribution of the parameters and the observed data, as in \eqref{joint}.  In a similar strategy to that
of \citet{hernandez2015expectation},
 we factorized the  joint distribution of the parameters and the observed data
to only three factors as follows

$$p(y,X,\beta,\Gamma,\eta,\theta|Z)= L(\beta,\Gamma|X,Z,y)p(\beta|\eta)p(\Gamma|\theta)p(\eta)p(\theta)=\prod_{i=1}^3f_i(\beta,\Gamma,\eta,\theta),$$
where $f_1(\beta,\Gamma,\eta,\theta)=L(\beta,\Gamma|X,Z,y)$,
$f_2(\beta,\Gamma,\eta,\theta)=p(\beta|\eta)p(\Gamma|\theta)$ and
$f_3(\beta,\Gamma,\eta,\theta)=p(\eta)p(\theta)$.

{ To imply the EP algorithm for approximation of the posterior function \eqref{post}, one might consider the factorization \eqref{joint} to the likelihood function \eqref{like}, the priors \eqref{pbet} and \eqref{pgam}, and the hyper-priors \eqref{peta} and \eqref{ptet}. Because of the complexity of the structure of the likelihood function $f_1(\beta,\Gamma)=L(\beta,\Gamma|X,Z,y)$, it is impossible to compute the sufficient statistics of $f_1\tilde{Q}^{(-1)}$, as needed in the EP algorithm for updating the parameters of the $\tilde{f}_1$. Thus, implementation of the EP algorithm is intractable based on the full likelihood function \eqref{like}. So, we
propose a two-stage modification of the EP algorithm here, which uses the partial likelihoods in each stage instead of the full likelihood \eqref{like}.} This
proposed modification is based on the standard idea of replacing
covariates $X$ by their expectations conditional on the instruments,
as in the classical two-stage least squares (2SLS) method
\citep{anderson2005origins},
in which the covariates $X$ are first
regressed on the instruments $Z$ and the response is then regressed
on the first stage predictors. This method is also used by
\citet{MR3338502},
who proposed a two stage regularization method
for high-dimensional instrumental variables regression. {Indeed, the simplification is done by replacing the complex covariance matrix $\Omega(\beta)=B\Sigma B'$ with a diagonal matrix $\Omega'={\rm diag}(\sigma_0^2,\tau_0^2,\tau_0^2,\ldots,\tau_0^2)$.}

The structure of the two-stage EP is as follows:

\begin{itemize}

\item Stage I:
\begin{description}
\item I-(i): Consider regressing the covariates $X$ on the instruments $Z$, that is,
$X=Z\Gamma+\varepsilon$, and the partial likelihood of this model as
$$L_p(\Gamma|X,Z)=\prod_{i=1}^n{\cal N}_p(X_i; Z_i\Gamma,\tau_0^2I_p),$$
{where $L_p$ stands for the partial likelihood. }
 Also, consider the prior \eqref{pgam} and the hyper-prior \eqref{ptet}.

\item I-(ii): Factorize the joint distribution
\begin{equation}\label{joint1}
p(X,\Gamma,\theta|Z)= L_p(\Gamma|X,Z)p(\Gamma|\theta)p(\theta)=\prod_{i=1}^3f_i(\Gamma,\theta),
\end{equation}
where $f_1(\Gamma,\theta)=L_p(\Gamma|X,Z)$,
$f_2(\Gamma,\theta)=p(\Gamma|\theta)$ and
$f_3(\Gamma,\theta)=p(\theta)$.

\item I-(iii): Apply the EP algorithm to approximate the joint distribution in \eqref{joint1} by

\begin{equation}\label{qtilde1}
\tilde{p}(X,\Gamma,\theta|Z)=\prod_{i=1}^3\tilde{f}_i(\Gamma,\theta)=\tilde{Q}_1(\Gamma,\theta),
\end{equation}
where
$$\tilde{f}_1(\Gamma,\theta)=\prod_{\ell=1}^{pq}{\cal N}(\gamma_{\ell}; \mu_{1\ell},\omega_{1\ell}),
$$
$$\tilde{f}_2(\Gamma,\theta)=\prod_{\ell=1}^{pq}{\cal N}(\gamma_{\ell}; \mu_{2\ell},\omega_{2\ell}){\rm Ber}(\theta_{\ell};\sigma(\pi_{2\ell})),
$$
$$\tilde{f}_3(\Gamma,\theta)=\prod_{\ell=1}^{pq}{\rm Ber}(\theta_{\ell};\sigma(\pi_{3\ell})),
$$
{ in which $\mu_{1\ell},\omega_{1\ell},\mu_{2\ell},\omega_{2\ell}, \pi_{2\ell} $ and $\pi_{3\ell}, \; \ell = 1,\ldots, pq,$ are parameters to be estimated,}
and $\sigma(x)=(1-e^{-x})^{-1}$ is the sigmoid function which
guarantees the success probability of the Bernoulli distributions to
be always in $(0,1)$. Continue the EP algorithm until convergence.
{The estimate of $\gamma$ is then obtained by the mean
of the approximated posterior, that is
$$\hat{\gamma}=\left(\frac{1}{\omega_1}+\frac{1}{\omega_2}\right)^{-1}\left(\frac{\mu_1}{\omega_1}+\frac{\mu_2}{\omega_2}\right).$$
Then, compute the predicted covariate $\hat{X}=Z\hat{\Gamma}$, in
which $\hat{\gamma}={\rm vec}(\hat{\Gamma})$. }
\end{description}

\item Stage II:
\begin{description}
\item II-(i): Consider regressing the responses $y$ on the predicted covariates $\hat{X}$ from Stage I, that is,
$y=\hat{X}\beta+u$, and the partial likelihood of this model as
$$L_p(\beta|\hat{X},y)=\prod_{i=1}^n{\cal N}(y_i; \hat{X}_i\beta,\sigma_0^2).$$
Also, consider the prior \eqref{pbet} and the hyper-prior \eqref{peta}.

\item II-(ii): Factorize the joint distribution
\begin{equation}\label{joint2}
p(y,\beta,\eta|\hat{X})= L_p(\beta|\hat{X},y)p(\beta|\eta)p(\eta)=\prod_{i=1}^3g_i(\beta,\eta),
\end{equation}
where $g_1(\beta,\eta)=L_p(\beta|\hat{X},y)$,
$g_2(\beta,\eta)=p(\beta|\eta)$ and
$g_3(\beta,\eta)=p(\eta)$.

\item II-(iii): Apply the EP algorithm to approximate the joint distribution in \eqref{joint2} by

\begin{equation}\label{qtilde2}
\tilde{p}(y,\beta,\eta|\hat{X})=\prod_{i=1}^3\tilde{g}_i(\beta,\eta)=\tilde{Q}_2(\beta,\eta),
\end{equation}
where
$$\tilde{g}_1(\beta,\eta)=\prod_{j=1}^p{\cal N}(\beta_j; m_{1j},\nu_{1j}),
$$
$$\tilde{g}_2(\beta,\eta)=\prod_{j=1}^p{\cal N}(\beta_j; m_{2j},\nu_{2j}){\rm Ber}(\eta_j;\sigma(p_{2j})),
$$
$$\tilde{g}_3(\beta,\eta)=\prod_{j=1}^p{\rm Ber}(\eta_j;\sigma(p_{3j})),
$$
{ in which $m_{1j},\nu_{1j},m_{2j},\nu_{2j}, p_{2j} $ and $p_{3j}, \; j = 1,\ldots, p,$ are parameters to be estimated.}
Continue the EP algorithm until convergence.

{The estimate of $\beta$ is then given by the mean of the approximated posterior, that is
$$\hat{\beta}=\left({\frac{1}{\nu_1}+\frac{1}{\nu_2}}\right)^{-1}\left(\frac{m_1}{\nu_1}+\frac{m_2}{\nu_2}\right).$$
To obtain final sparse estimates of $\beta$ and $\gamma$, we let $$\hat{\beta}_j=0, \quad \mbox{if}\quad \sigma(-p_{2j}-p_{3j})>{\rm Q}_{p_0}(\sigma(-p_{2}-p_{3})),$$
and
$$\hat{\gamma}_j=0, \quad \mbox{if}\quad \sigma(-\pi_{2j}-\pi_{3j})>{\rm Q}_{\pi_0}(\sigma(-\pi_{2}-\pi_{3})),$$
where ${\rm Q}_t(v)$ is the $t$th quantile of the vector $v$.
}
\end{description}
\item {Finally, a post-estimation method is performed to obtain the final estimators using the ridge regression technique applied to the selected variables. }
\end{itemize}

{ The details of the algorithm are given in the Appendix.}

{\subsection{Initializing the model}

In practice, the parameters $\sigma_0^2$ and $\tau_0^2$ and the hyper-parameters $p_0$, $\pi_0$, $\nu_0$ and $\omega_0$ are unknown.
The model can be initialized using one of the following strategies:
\begin{description}
\item Strategy I: Initialize the model by first applying the 2-stage method of \citet{MR3338502}
to the data set along
with a model selection criterion such as AIC or BIC to select the optimal model and obtain $\hat{\beta}^{\rm init}$ and $\hat{\Gamma}^{\rm init}$, and then we initialize the model as follows
$$\hat{p}_0={\rm df}_1/p,\quad {\rm df}_1=\#\{j;\;\hat{\beta}_j^{\rm init}\neq 0,\; 1\leq j\leq p\}$$
$$\hat{\pi}_0={\rm df}_2/(pq),\quad {\rm df}_2= \#\{j;\;\hat{\gamma}_j^{\rm init}\neq 0,\; 1\leq j\leq pq\}$$
\begin{equation}\label{sigh}\hat{\sigma}_0^2=||y-\hat{X}\hat{\beta}^{\rm init}||_2^2/{\rm df}*,\quad {\rm df}*=\left\{\begin{array}{lr}
n-{\rm df}_1, & {\rm if}\; {\rm df}_1<n\\
n/2, & {\rm otherwise}\end{array}\right.,
\end{equation}
\begin{equation}\label{tauh}
\hat{\tau}_0^2=||X-Z\hat{\Gamma}^{\rm init}||_F^2/n,
\end{equation}
{ and
\begin{equation}\label{nuom}
\hat{\nu}_0=\sum_{j=1}^p(\hat{\beta}_j^{\rm init})^2/{\rm df}_1,\quad \hat{\omega}_0=\sum_{j=1}^{pq}(\hat{\gamma}_j^{\rm init})^2/{\rm df}_2,
\end{equation}
}
where $\#A$ stands for the cardinality of the set $A$, $\hat{\gamma}^{\rm init}={\rm vec}(\hat{\Gamma}^{\rm init})$, $||v||_2^2=v'v$
is the squared norm of vector $v$ and $||B||_F^2$ is the squared Frobenius
norm of matrix $B$. This strategy is used in the simulation study, in Section 4.

\item Strategy II: Initialize the model as in Strategy I, let
$\hat{\sigma}_0^2$, $\hat{\tau}_0^2$, {$\hat{\nu}_0$ and $\hat{\omega}_0$ be as in \eqref{sigh} to
\eqref{nuom}}, respectively. Seek for the
optimal values of $p_0$ and $\pi_0$ through a grid of values, based
on the cross-validation, AIC or BIC criteria. This strategy is used in the real data analysis in Section 4.
\end{description}}

\section{Simulation study}

{In this section, a Mont\'{e} Carlo simulation study is conducted, in order to
examine the performance of the proposed method. For this purpose, the IV model with {$p=300$, $q=400$, $n=50$,
and the following parameters is considered
$$\beta=(\mathbf{1}_{7}',\mathbf{0}_{285}',-0.5\cdot \mathbf{1}_{8}')',$$
$$\gamma=(0.01\cdot\mathbf{1}_{300},\mathbf{0}_{118800},-0.005\cdot \mathbf{1}_{900}),$$
where $\mathbf{1}_p$ and $\mathbf{0}_p$ stand for the vector of 1s and 0s with length $p$, respectively,
which means that
$$\Gamma=\left(\begin{array}{c}
0.01\cdot\mathbf{1}_{1\times 300}\\
\mathbf{0}_{396\times 300}\\
-0.005\cdot\mathbf{1}_{3\times 300}
\end{array}\right),$$
where $\mathbf{1}_{p\times q}$ and $\mathbf{0}_{p\times q}$ stand for the $p\times q$ matrices of 1s and 0s, respectively. The vector $\beta$ is set such that 5\% of its elements are non-zero, while this ratio is equal 1\% for the vector $\gamma$.
}
{
The number of repeated simulated data sets for the Monte Carlo simulation study is $N=10^3$ iterations. In each iteration:
\begin{enumerate}
\item  {The genotype data, $Z_{ij}$, is generated from Bernoulli distribution with a success probability of $r_{ij}$, for $i=1,\ldots n$, $j=1,\ldots,q$, where
$r_{ij}$s are generated from Beta distribution with parameters 3 and 7 (with mean 0.3 and standard deviation 0.138). This model tries to simulate a complicate phenomenon similar to the real genotype data which depends on Minor allele frequency (MAF) and Hardy-Weinberg Equilibrium.}
\item $X_{ij}$ is generated from
$N(0.1+Z_i\Gamma_j,0.1)$, for $i=1,\ldots,n$, $j=1,\ldots,p$,
\item $y_i$ is generated from $N(1+X_i\beta,0.5)$, for $i=1,\ldots,n$.
\end{enumerate}
}

The two-stage EP algorithm is applied in each iteration to estimate the parameters.
As a result of the simulation study, the  false negative rate and the false positive rate,
defined as follows, are computed for estimation of $\beta$ and $\Gamma$,
$${\rm  FNR}_{\beta}=\frac{\#\{j;\;1\leq j\leq p,\; \beta_j\neq 0,\; \hat{\beta}_j=0\}}{\#\{j;\;1\leq j\leq p,\; {\beta}_j\neq 0\}},$$
$${\rm  FPR}_{\beta}=\frac{\#\{j;\;1\leq j\leq p,\; {\beta}_j= 0,\; \hat{\beta}_j\neq 0\}}{\#\{j;\;1\leq j\leq p,\; {\beta}_j= 0\}},$$
$${\rm  FNR}_{\Gamma}=\frac{\#\{\ell;\;1\leq \ell\leq pq,\; \gamma_{\ell}\neq 0,\; \hat{\gamma}_{\ell}=0\}}{\#\{j;\;1\leq {\ell}\leq pq,\; {\gamma}_{\ell}\neq 0\}}$$
and
$${\rm  FPR}_{\Gamma}=\frac{\#\{{\ell};\;1\leq {\ell}\leq pq,\; {\gamma}_{\ell}= 0,\; \hat{\gamma}_{\ell}\neq 0\}}{\#\{{\ell};\;1\leq {\ell}\leq pq,\; {\gamma}_{\ell}= 0\}},$$
where $\#A$ stands for the {cardinality} of the set $A$.

{Furthermore, 3-fold cross-validation (CV) criterion
$${\rm CV}=\frac{1}{3}\sum_{j=1}^3\sum_{i\in F_j}\left({y}_i-Z_i \hat{\Gamma}_{(-i)}\hat{\beta}_{(-i)}\right)^2,$$
are computed, where $\{F_1,F_2,F_3\}$ is a partition of $\{1,\ldots,n\}$.
}

{The two-stage EP (2S.EP) method is compared with its
two frequentist competitors proposed by \citet{MR3338502},
which are two-stage sparse IV model based on the LASSO (2S.LASSO) and SCAD
(2S.SCAD) penalties. Figure \ref{simul} shows the box-plots of
${\rm  FNR}_{\beta}$, ${\rm  FPR}_{\beta}$, ${\rm
FNR}_{\Gamma}$, ${\rm  FPR}_{\Gamma}$, ${\rm CV}$ and the computation time
for 2S.EP, 2S.LASSO and 2S.SCAD. As one
can see from Figure \ref{simul}, the 2S.EP method performs better than 2S.LASSO and 2S.SCAD, in detecting the effective and non-effective covariates (In terms of FPR and FNR), while it has a poor prediction performance and more computation time relative to the frequentist methods  2S.LASSO and 2S.SCAD. }

{\begin{figure}
\centering
\includegraphics[scale=0.6]{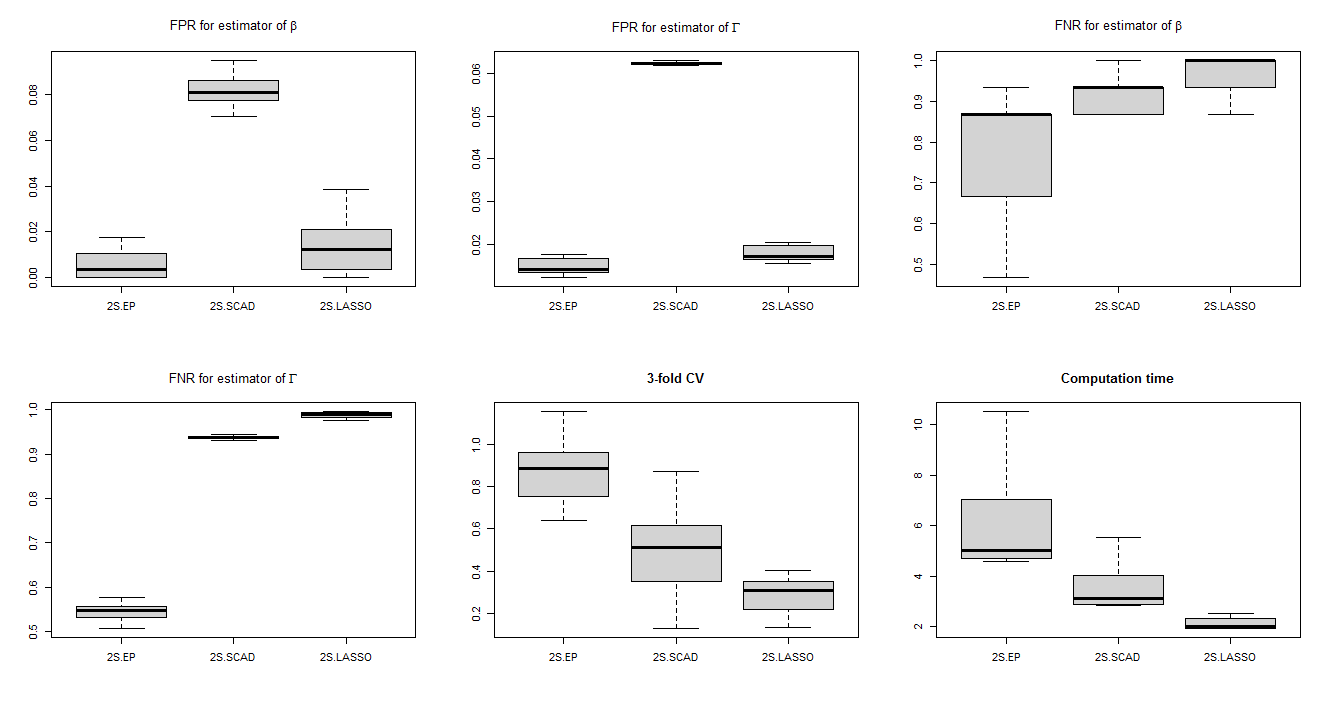}
\caption{The results of the simulation study, for comparison of
the 2S.EP method with 2S.LASSO and 2S.SCAD methods, with $N=10^3$ iterations. The  relative errors are computed
as the result of dividing the corresponding error of 2S.LASSO and 2S.SCAD methods by that of 2S.EP. }\label{simul}
\end{figure}}

\section{Analysis of mouse obesity data}

After the individuals, SNPs, and genes with a missing rate
greater than 0.1 were removed, the remaining missing genotype
and gene expression data were imputed using the linkage based imputation method \citep{xu2015linkage}
and nearest neighbor
averaging \citep{troyanskaya2001missing},
respectively. Merging the
genotype, gene expression, and clinical data yielded a complete
data-set with $q = 2654$ SNPs and $23184$ genes on $n = 290$ mice. To enhance the interpretability and
stability of the results, we focus on the $p = 3041$ genes that
have standard deviation of gene expression levels
greater than $0.1$. The latter criterion is reasonable because
gene expressions of too small variation are typically not of biological
interest and suggest that the genetic perturbations may
not be sufficiently strong for the genetic variants to be used as
instruments.

\begin{figure}
\centering
\includegraphics[scale=0.6]{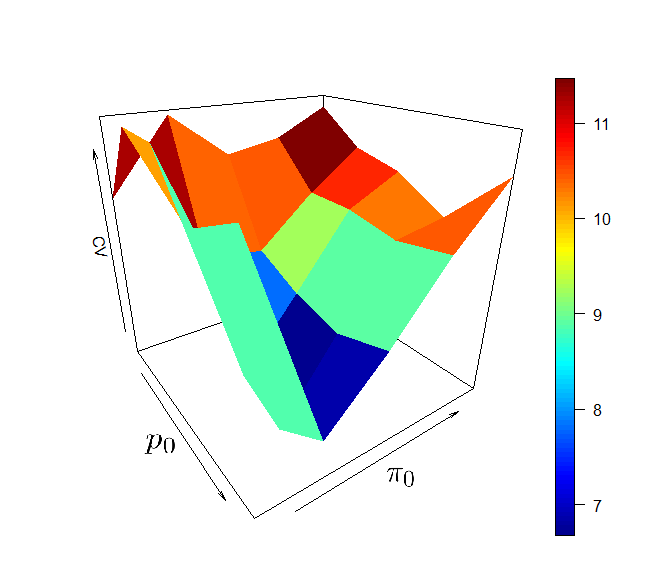}
\caption{The 3-fold cross-validation as a function of $p_0$ and $\pi_0$.}\label{cvfig}
\end{figure}

\begin{figure}
\centering
\includegraphics[scale=0.4]{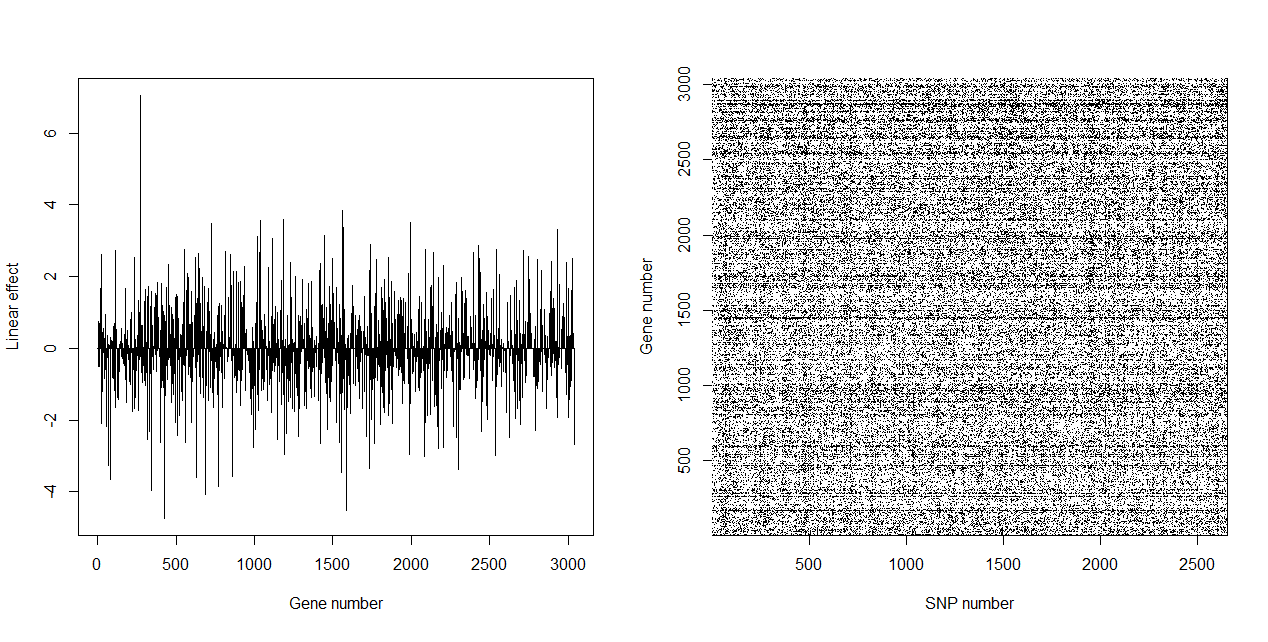}\\
\includegraphics[scale=0.4]{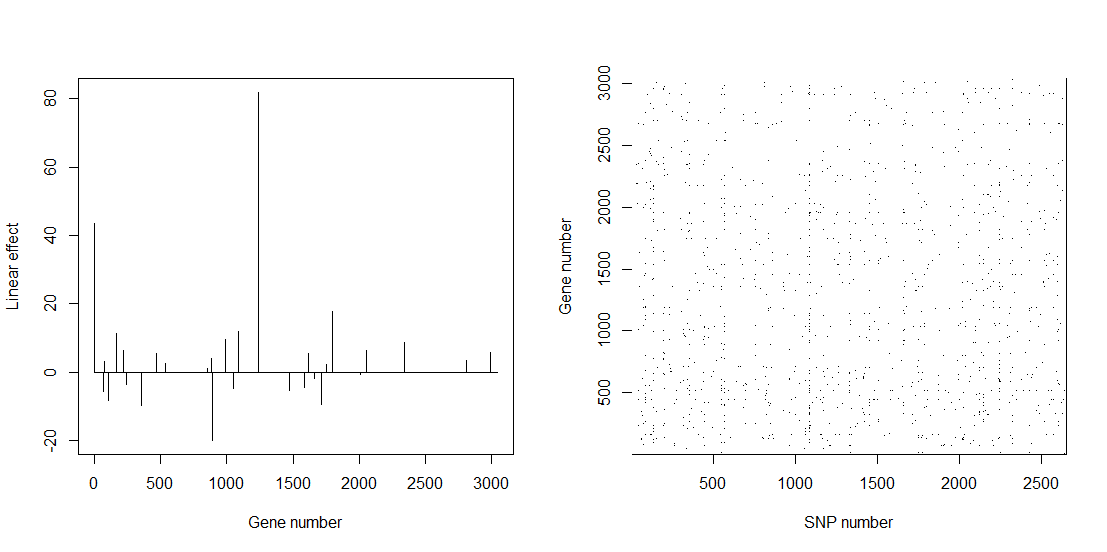}\\
\includegraphics[scale=0.4]{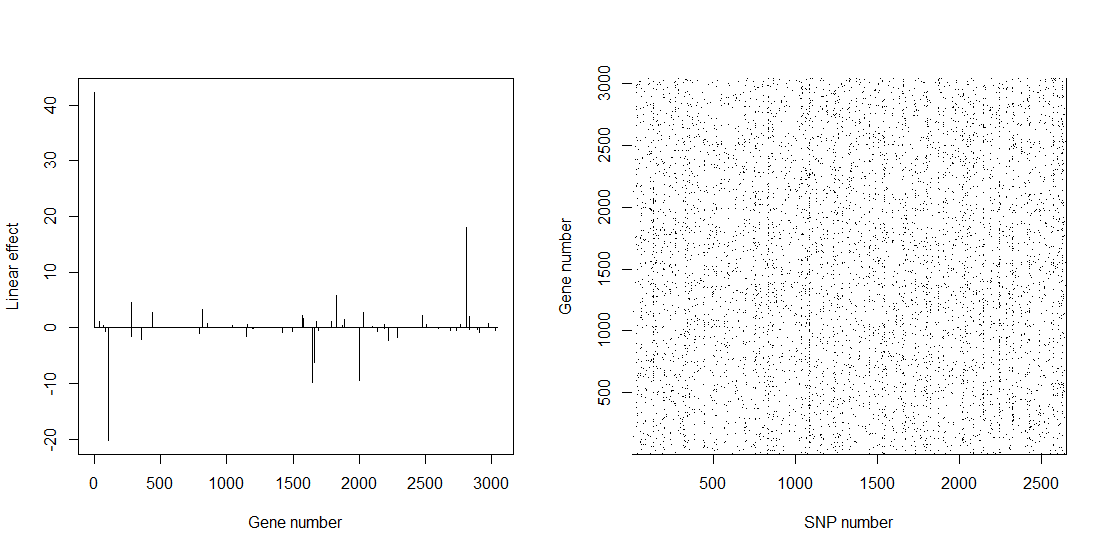}
\caption{Estimation of $\beta$ (left) and $\Gamma$ (right) for two stage EP
LASSO (second row) and SCAD (third row) methods.}\label{betfig}
\end{figure}

Our goal is to jointly analyze the genotype, gene expression,
and clinical data to identify important genes related to body
weight.

The two-stage EP algorithm (2S.EP), proposed in the previous section, as well as the two-stage LASSO (2S.LASSO) and SCAD (2S.SCAD) methods, proposed by \citet{MR3338502},
are applied to the mouse obesity data-set. } { For the two-stage EP algorithm, Strategy II is used to initialize the hyper-parameters, using the 3-fold cross-validation as the criterion and the maximum errors for both stages was $10^{-4}$.  Figure \ref{cvfig} shows the 3D plot of 3-fold cross-validation as a function of $p_0$ and $\pi_0$. The values of $p_0$ and $\pi_0$ are selected from the { sequence from 0.1 to 0.9 with steps of 0.2}. The optimal values are {$p_0=0.7$ and $\pi_0=0.3$.}

{Figure \ref{betfig} shows the sparse estimates of the coefficients $\beta$ (left) and $\Gamma$ (right) for the mouse obesity data-set, based on the two stage EP (up) LASSO (middle) and SCAD (down) methods.  The values of the non-zero effects of the genes (covariates) on the response can be seen from the left panel of Figure \ref{betfig}, while in the right panel, a 2D sparse plot of the estimate of the coefficient matrix $\hat{\Gamma}$ is shown. The black dots and lines represent the non-zero estimates. {The exact estimates as well as the effective genes and SNPs are available at \href{https://github.com/mortamini/2Stage-Sparse-IVR}{https://github.com/mortamini/2Stage-Sparse-IVR}.}}

{Based on the obtained estimates, the coefficient of determination for prediction of the response $y$ given $Z$, $R^2_{{ y|\hat{X}}}$, the 3-fold cross-validation, CV, and the Bayesian Information Criterion,  ${\rm BIC}_{{ y|\hat{X}}}$, are given in Table \ref{eval}. As one can see from Table \ref{eval}, the 2S.EP method is preferred based on the BIC criterion, while the 2S.LASSO method has a lower CV. }

\begin{table}
\centering{
\caption{The evaluation criteria for the three methods.}\label{eval}
\begin{tabular}{ c c c c }
\hline\hline
method & $R^2_{{ y|\hat{X}}}$ & CV & ${\rm BIC}_{{ y|\hat{X}}}$\\
2S.EP &   {\bf 0.99}    &  5.27    &  {\bf 17079}   \\
2S.LASSO &   0.61    &    {\bf 4.19}  &   18161  \\
2S.SCAD &   0.84    &   4.67   &   17898  \\
\hline\hline
\end{tabular}}
\end{table}

\section{Concluding remarks}

The causal inference using the Bayes method and based on the
sparsity-enforcing priors is considered in this paper and the EP
method is used for approximation of the posterior distribution. { An
advantage of using the Bayesian causal inference is that the posterior distribution of the estimators are obtained. Also, the results of the simulation study shows that the 2S-EP method  performs better than 2S.LASSO and 2S.SCAD, in detecting the effective and non-effective covariates.}

{The R functions to implement the proposed methods as
well as other 2-stage sparse frequentist competitors are available
at
\href{https://github.com/mortamini/2Stage-Sparse-IVR}{https://github.com/mortamini/2Stage-Sparse-IVR}.}
{ The post estimation is also considered in the prepared
functions, which is re-estimation of the model parameters after
removing the ineffective covariates from the model, using
frequentist ordinary or Ridge models. The execution time of the
codes should be improved by calling C routines within the R codes
for the EP algorithm in each stage, and by using parallel
programming.}

{It is worth noting that the proposed results of this
paper could be improved by further cross-validation over all
parameters of the model, which was ignored for the matter of time.}

\section*{Acknowledgements}
The author would like to thank the anonymous referee for his/her valuable comments and suggestions on an earlier version of this article, which significantly improved the paper. This research was partially support by Iranian National Science Foundation under the grant number 99009577.

{\section*{Appendix (details of the algorithm)}
Using the product rule of the normal and Bernoulli densities, and by considering the normalizing constants, the approximated posterior distributions obtained from \eqref{qtilde1} and \eqref{qtilde1} are
\begin{eqnarray}
\tilde{p}(X,\Gamma,\theta|Z)&=&\prod_{\ell=1}^{pq}{\cal N}(\gamma_{\ell}; \xi_{\gamma_{\ell}},s^2_{\gamma_{\ell}}){\rm Ber}(\theta_{\ell};\sigma(u_{\theta_{\ell}})),\label{ptilde1}
\end{eqnarray}
and
\begin{eqnarray}
\tilde{p}(y,\beta,\eta|\hat{X})&=&\prod_{j=1}^p{\cal N}(\beta_j; \xi_{\beta_j},s^2_{\beta_j}){\rm Ber}(\eta_j;\sigma(u_{\eta_j})),\label{ptilde2}
\end{eqnarray}
respectively, where, for $j=1,\ldots,p$ and $\ell=1,\ldots,pq$,
$$\xi_{\beta_j}=[m_{1j}(\nu_{1j})^{-1}+m_{2j}(\nu_{2j})^{-1}]s^2_{\beta_j},$$
$$s^2_{\beta_j}=[(\nu_{1j})^{-1}+(\nu_{2j})^{-1}]^{-1},$$
$$\xi_{\gamma_{\ell}}=[\mu_{1\ell}(\omega_{1\ell})^{-1}+\mu_{2\ell}(\omega_{2\ell})^{-1}]s^2_{\gamma_{\ell}},$$
$$s^2_{\gamma_{\ell}}=[(\omega_{1\ell})^{-1}+(\omega_{2\ell})^{-1}]^{-1},$$
$$u_{\eta_j}=p_{2j}+p_{3j},\quad u_{\theta_{\ell}}=\pi_{2\ell}+\pi_{3\ell}.$$

Thus, for $j=1,\ldots,p$ and $\ell=1,\ldots,pq$, final non-sparse estimates of $\beta_j$ and $\gamma_{\ell}$ are $\xi_{\beta_j}$ and $\xi_{\gamma_{\ell}}$, respectively. For the purpose of variable selection and obtaining the sparse estimates, one can let $\hat{\beta}_j=0$, if $\sigma(u_{\eta_j})<\alpha_1$, and $\hat{\gamma}_{\ell}=0$, if $\sigma(u_{\theta_{\ell}})<\alpha_2$, for suitable threshold values, $\alpha_i\in(0,1),\; i=1,2.$

With an adapted approach to that used in \citet{hernandez2015expectation},
one can show that, in the first step of both EP algorithms, in two stages, the parameters of $\tilde{f}_3$ and $\tilde{g}_3$ are updated and do not change in the next steps, as follows
$$p_{3j}=\sigma^{-1}(p_0),\quad \pi_{3\ell}=\sigma^{-1}(\pi_0),\quad j=1,\ldots,p, \quad
\ell=1,\ldots,pq.$$

Furthermore, in step $t+1$, $t=0,\ldots,T-1$, the parameters of $\tilde{f}_2$ and $\tilde{g}_2$ are updated in step $t+1$, for $j=1,\ldots,p$ and ${\ell}=1,\ldots,pq$, as
$$\nu_{2j}^{(t+1)}=((a_j^{(t+1)})^2-b_j^{(t+1)})^{-1}-\nu_{1j}^{(t)}, $$
$$\omega_{2\ell}^{(t+1)}=((c_{\ell}^{(t+1)})^2-d_{\ell}^{(t+1)})^{-1}-\omega_{1{\ell}}^{(t)}, $$
$$m^{(t+1)}_{2j}=m^{(t)}_{1j}-a_j^{(t+1)}(\nu_{2j}^{(t+1)}+\nu_{1j}^{(t)}), $$
$$\mu^{(t+1)}_{2{\ell}}=\mu^{(t)}_{1{\ell}}-c_{\ell}^{(t+1)}(\omega_{2{\ell}}^{(t+1)}+\omega_{1{\ell}}^{(t)}), $$
$$p_{2j}^{(t+1)}=\frac{1}{2}\log(\nu_{1j}^{(t)})-\frac{1}{2}\log(\nu_{1j}^{(t)}+\nu_0)+\frac{1}{2}(m_{1j}^{(t)})^2\left[(\nu_{1j}^{(t)})^{-1}-
(\nu_{1j}^{(t)}+\nu_0)^{-1}\right],$$
$$\pi_{2{\ell}}^{(t+1)}=\frac{1}{2}\log(\omega_{1{\ell}}^{(t)})-\frac{1}{2}\log(\omega_{1{\ell}}^{(t)}+\omega_0)+\frac{1}{2}(\mu_{1{\ell}}^{(t)})^2\left[(\omega_{1{\ell}}^{(t)})^{-1}-
(\omega_{1{\ell}}^{(t)}+\omega_0)^{-1}\right],$$
where for $j=1,\ldots,p$ and ${\ell}=1,\ldots,pq$
$$a_j^{(t+1)}=\sigma(p_{2j}^{(t+1)}+p_{3j})\frac{m_{1j}^{(t)}}{\nu_{1j}^{(t)}+\nu_0}+\sigma(-p_{2j}^{(t+1)}-p_{3j})\frac{m_{1j}^{(t)}}{\nu_{1j}^{(t)}},$$
$$c_{\ell}^{(t+1)}=\sigma(\pi_{2{\ell}}^{(t+1)}+\pi_{3{\ell}})\frac{\mu_{1{\ell}}^{(t)}}{\omega_{1{\ell}}^{(t)}+\omega_0}+\sigma(-\pi_{2{\ell}}^{(t+1)}-\pi_{3{\ell}})\frac{\mu_{1{\ell}}^{(t)}}{\omega_{1{\ell}}^{(t)}},$$
$$b_j^{(t+1)}=\sigma(p_{2j}^{(t+1)}+p_{3j})\frac{(m_{1j}^{(t)})^2-\nu_{1j}^{(t)}-\nu_0}{(\nu_{1j}^{(t)}+\nu_0)^2}+\sigma(-p_{2j}^{(t+1)}-p_{3j})\left[(m_{1j}^{(t)})^2(\nu_{1j}^{(t)})^{-2}-(\nu_{1j}^{(t)})^{-1}\right],$$
$$d_{\ell}^{(t+1)}=\sigma(\pi_{2{\ell}}^{(t+1)}+\pi_{3{\ell}})\frac{(\mu_{1{\ell}}^{(t)})^2-\omega_{1{\ell}}^{(t)}-\omega_0}{(\omega_{1{\ell}}^{(t)}+\omega_0)^2}+\sigma(-\pi_{2{\ell}}^{(t+1)}-\pi_{3{\ell}})\left[(\mu_{1{\ell}}^{(t)})^2(\omega_{1{\ell}}^{(t)})^{-2}-(\omega_{1{\ell}}^{(t)})^{-1}\right],$$

To avoid the updated values of the parameters $\nu_{2j}$ and $\omega_{2\ell}$ to be negative, \citet{hernandez2015expectation}
suggest to update the parameters of $\tilde{f}_2$ and $\tilde{g}_2$ by minimizing
$${\rm KL}(f_2 Q_1^{(-2)}|| \tilde{f}_2Q_1^{(-2)})\quad \mbox{and}\quad {\rm KL}(g_2 Q_2^{(-2)}|| \tilde{g}_2Q_2^{(-2)}),$$ under the constraint $\nu_{2j}\geq 0$, $\omega_{2\ell}\geq 0$, respectively, and proved that this will result if infinite optimal value of $\nu_{2j}$ and $\omega_{2\ell}$. Thus,
whenever each of these parameters get negative, we simply replace them by a large positive constant.

The update of the parameters of $\tilde{g}_1$ is again similar to
that of \citet{hernandez2015expectation},
while that of
$\tilde{f}_1$, is somehow different from that of
\citet{hernandez2015expectation},
partly because of the $p$-variate
normal density component in $p(X|Z,\Gamma)$.   For $t=0,\ldots,T-1$,
letting ${\cal V}_2^{(t)}$ and ${\cal W}_2^{(t)}$ be the diagonal
matrices with diagonal elements
$(\nu_{21}^{(t)},\ldots,\nu_{2p}^{(t)})$ and
$(\omega_{21}^{(t)},\ldots,\omega_{2(pq)}^{(t)})$, respectively, the
updated parameters of $\tilde{f}_1$ and $\tilde{g}_1$ in step $t+1$
of the EP algorithms, for $j=1,\ldots,p$ and ${\ell}=1,\ldots,pq$,
are
$$\nu_{1j}^{(t+1)}=\left[(V_{jj}^{(t+1)})^{-1}-(\nu_{2j}^{(t)})^{-1}\right]^{-1},$$
$$\omega_{1j}^{(t+1)}=\left[(W_{\ell\ell}^{(t+1)})^{-1}-(\omega_{2{\ell}}^{(t)})^{-1}\right]^{-1},$$
$$m_{1j}^{(t+1)}=\left[{\cal M}_j^{(t+1)}(V_{jj}^{(t+1)})^{-1}-m_{2j}^{(t)}(\nu_{2j}^{(t)})^{-1}\right]\nu_{1j}^{(t+1)},$$
$$\mu_{1{\ell}}^{(t+1)}=\left[{\cal N}_{\ell}^{(t+1)}(W_{{\ell}{\ell}}^{(t+1)})^{-1}-\mu_{2{\ell}}^{(t)}(\omega_{2{\ell}}^{(t)})^{-1}\right]\omega_{1{\ell}}^{(t+1)},$$
where
$$V^{(t+1)}=[({\cal V}_2^{(t)})^{-1}+\sigma_0^{-2}\hat{X}'\hat{X}]^{-1}={\cal V}_2^{(t)}-{\cal V}_2^{(t)}\hat{X}'[\sigma_0^2 I_n+\hat{X}{\cal V}_2^{(t)}\hat{X}']^{-1}\hat{X}{\cal V}_2^{(t)},$$
\begin{align*}
W^{(t+1)}&=[({\cal W}_2^{(t)})^{-1}+\tau_0^{-2}I_{p}\otimes(Z'Z)]^{-1}\\
&={\cal W}_2^{(t)}-{\cal W}_2^{(t)}Z'\otimes I_p[\tau_0^2 I_{np}+(Z\otimes I_p){\cal W}_2^{(t)}(Z'\otimes I_p)]^{-1}Z\otimes I_p{\cal W}_2^{(t)},
\end{align*}
$${\cal M}^{(t+1)}=V^{(t+1)}[({\cal V}_2^{(t)})^{-1}m_{2}^{(t)}+\sigma_0^{-2}\hat{X}'y],$$
$${\cal N}^{(t+1)}=W^{(t+1)}[({\cal W}_2^{(t)})^{-1}\mu_{2}^{(t)}+\tau_0^{-2}Z'X],$$
and $\otimes$ stands for the Kronecker product.

In many problems, { especially} for the genetic association problems, $p$ and $q$ are large values, and thus computation of the
updated matrix $W^{(t+1)}$ and vector ${\cal M}^{(t+1)}$ in step $t+1$ of the EP algorithm needs huge amount of memory. To reduce the used
memory for each computation and provide suitable formulas for parallel computations, one can use the fact that
${\cal W}_2^{(t)}$, $Z\otimes I_p$, $Z'\otimes I_p$ and $I_{np}$ are block diagonal matrices and decompose the computations into the following sub-computations
$$W^{(t+1)}_{(j)}={{\cal W}_2}_{(j)}^{(t)}-{{\cal W}_2}_{(j)}^{(t)}Z'[\tau_0^2 I_{n}+Z{{\cal W}_2}_{(j)}^{(t)}Z')]^{-1}Z{{\cal W}_2}_{(j)}^{(t)},$$
$${\cal N}^{(t+1)}_{(j)}=W^{(t+1)}_{(j)}[({{\cal W}_2}_{(j)}^{(t)})^{-1}{\mu_{2}}_{(j)}^{(t)}+\tau_0^{-2}C_{j}],$$
for $j=1,\ldots,p$, where $A_{(j)}$ stand for the $j$th diagonal block of the diagonal matrix $A$, ${\mu_{2}^{(t)}}=({\mu_{2}}_{(1)}^{(t)},\ldots,{\mu_{2}}_{(p)}^{(t)})$ and $C_{j}$ is the $j$th row of the matrix $C=Z'X$.}



\end{document}